\documentclass[a4paper,12pt]{article}

\usepackage[utf8]{inputenc}
\usepackage{xcolor}
\usepackage{url}

\usepackage{hyperref}
\usepackage{graphicx}
\usepackage{subcaption}
\usepackage{tabularx}
\usepackage{booktabs}
\usepackage{soul}

\newcommand{\tabhead}[1]{\textbf{#1}}

\newcommand{\mean}[1]{\langle#1\rangle}
\newcommand{\E}{\mathbf{E}}

\newcommand{\avgdeg}{\mean{\mathrm{deg}}}

\usepackage{geometry}
\title{Epidemics in a Synthetic Urban Population with Multiple Levels of Mixing}
\author{Alessandro Celestini\thanks{Institute for Applied Computing, CNR, Rome, Italy.
 (alessandro.celestini@cnr.it, stefano.guarino@cnr.it, enrico.mastrostefano@cnr.it)} \and Francesca Colaiori\thanks{Institute for Complex Systems, CNR, Rome, Italy 
 (francesca.colaiori@cnr.it)} 
 \thanks{Department of Physics, Sapienza University, Rome, Italy} 
 \and Stefano Guarino$^*$ \and Enrico Mastrostefano$^*$  \and  Lena Rebecca Zastrow$^*$ }

\date{}

\begin{document}

\maketitle             

\begin{abstract}
Network--based epidemic models that account for heterogeneous contact patterns are extensively used to predict and control the diffusion of infectious diseases. We use census and survey data to reconstruct a geo--referenced and age--stratified synthetic urban population connected by stable social relations. We consider two kinds of interactions, distinguishing daily (household) contacts from other frequent contacts. Moreover, we allow any couple of individuals to have rare fortuitous interactions. We simulate the epidemic diffusion on a synthetic urban network for a typical medium-size Italian city and characterize the outbreak speed, pervasiveness, and predictability in terms of the socio--demographic and geographic features of the host population. 
Introducing age--structured contact patterns results in faster and more pervasive outbreaks, while assuming that the interaction frequency decays with distance has only negligible effects. 
Preliminary evidence shows the existence of patterns of hierarchical spatial diffusion in urban areas, with two regimes for epidemic spread in low- and high-density regions.

\end{abstract}

\section{Introduction}\label{sec:intro}
The prediction and control of infectious diseases have an enormous impact on public health, the economy, and society. 
Mathematical modeling and computer simulations provide powerful tools to understand the dynamics of epidemic outbreaks and design strategies to control and mitigate them. 
The modeling of epidemic contagion has evolved from simple compartmental schemes to sophisticated data--informed models using synthetic populations whose demographics are statistically indistinguishable from the census data. 
As in other fields, this evolution towards more realistic models involves a trade--off between simplicity and accuracy of the predictions.    

The classical compartmental models of disease propagation~\cite{Kermack1991,hethcote2000mathematics} describe the system at the population--level by grouping individuals in sub--populations according to their health states relevant for transmission and tracking changes in the sizes of the sub--populations, without specifying which individuals are involved.
The movements in and out of the compartments are governed by constant transition rates in ordinary differential equations, corresponding to exponentially distributed waiting times in the compartments.
This approach assumes a fully--mixed population where an infective individual is equally likely to spread the disease to any other individual, and it is equivalent to a mean--field.

More recent works focused on incorporating heterogeneous contact patterns in the models.
In individual--based models, the population--level behavior emerges from the microscopic interactions between agents that can carry a set of attributes such as age, spatial location, and social behavior.
Socio--demographic attributes collected from census data and surveys, or integrated with mobile, traffic or wearable sensor data and online tools~\cite{Mossong-2008,cattuto2010dynamics,mistry2020inferring}, allow the construction of realistic contact matrices describing the mixing patterns in typical social settings (e.g., households, schools, workplaces)~\cite{Mossong-2008,eubank2004modelling,merler2009role,chang2021mobility}.
Meta--population models built using data from inter--population mobility~\cite{hufnagel2004forecast,colizza2006role,chang2021mobility,tizzoni2014use} have permitted remarkable progress in describing how the disease travels from one city/nation to the other and estimating epidemic paths predictability~\cite{Colizza2007}. 
Compartmental models defined with stochastic and time--varying disease transmission rates \cite{cazelles2018accounting} can include changes in the transmission rate that might occur due to the implementation of control measures and variations due to unobserved processes.

By representing individuals in the population as vertices of a graph, network--based models aim to understand how the topological structure of the contact network affects the spreading dynamics upon them~\cite{gupta1989networks,newman2002spread,keeling2005networks,chakrabarti2008epidemic,pastor2015epidemic}.
Starting from actual epidemiological data, a recent body of research work \cite{charu2017human,grenfell2001travelling,li2018effect} tried to reconstruct the relationship between population heterogeneity and the spread of infectious diseases in a population or territory.
Due to the lack of granular disease data, a clear picture is still missing.
The influence of population density and distances, examined at different scales (world vs. states vs. cities), leads to different results on the main driver of diffusion. 

Ball and Neal~\cite{ball2002general,ball2008network} addressed the need to incorporate randomness in network--based epidemic models studying a stochastic Susceptible--Infected--Removed (SIR) model in a closed population where individuals make {\it local} contacts on a social network and {\it global} contacts with individuals chosen at random in the population.
This model is also suited to mimic multiple routes of transmission \cite{kiss2006effect}.

In this paper, we address similar questions for epidemics at the urban scale by simulating an agent--based SIR model in a synthetic urban population.
As in~\cite{ball2008network}, our model incorporates casual contacts on top of a structured set of social relations.
However, we use a data--informed model to obtain an age--structured population connected by a geographically referenced social network, and we allow for global casual contacts to be dependent on distance and/or on the age class of an agent. 
By encoding in our epidemic model the dependence of contact frequencies on age and distance, we can follow the geographical and age--stratified evolution of the outbreak under different configurations, without the need for high--resolution data about transportation, mobility, workplaces, or schools -- which, however, may be incorporated in the model in the future.
The goal is to understand how, in a realistic population bound by a data--driven social network, the combination of different levels of mixing and their possible dependence on socio--demographic and geographic features of the host population impact the epidemic diffusion process.

\section{Models, Materials and Methods}\label{sec:model}

We consider the \emph{territory} of the Municipality of Viterbo, Italy, represented as a rectangular bounding box, partitioned into a grid of $T$ square tiles of fixed side $l=500$ m.
We build a \emph{synthetic population} of $N\approx 60  K$ geo--referenced and age-stratified agents using density estimates provided by the WorldPop Project~\cite{Wordpop2020} and keep only tiles having at least 10 residents.
Each agent is assigned to an age group based on census data aggregated at the provincial level.
We consider four age--groups: \emph{children} (0 to 17), \emph{young} people (18 to 34), \emph{adults} (35 to 64), and \emph{elderly} (65+).
The $N$ agents form the vertex set $V$ of an urban social network (USN), i.e., an unweighted undirected graph $G=(V,E)$ obtained as the union of the \emph{household graph} $G_H=(V,E_H)$ and \emph{acquaintance graph} $G_A=(V,E_A)$, where $E=E_H\cup E_A$ and $E_H\cap E_A = \emptyset$.
The graph $G_H$ consists of a set of \emph{cliques}, one per household, inferred from census data under the assumption that: $(i)$ all members of a household live in the same tile; $(ii)$ children are younger than their parents; $(iii)$ partners have, on average, a similar age.
The algorithm used to group individuals into households is described in detail in~\cite{Inferring}.
The resulting household graph has a data--driven average degree $\nu$, which has been found empirically to be $\approx 2$.
To construct the acquaintance graph $G_A$, for each pair $(u,v)$, the edge $(u,v)$ is added to $E_A$ independently of all others with probability $\psi_{u,v}=\Pr[(u,v)\in E_A]\propto s_{g_u,g_v}\cdot d(u,v)^{-1} \cdot f_u \cdot f_v$, where: 
$s_{i,j}$ is the age-–based social mixing for groups $i$ and $j$, deduced from aggregated contact data~\cite{Mossong-2008} through the SOCRATES Data Tool~\cite{willem2020socrates};
$d(u,v)=\max\left\{\frac{l}{2},d^*(u,v)\right\}$ is the discretized distance between $u$ and $v$, where $d^*(u,v)$ is the distance between the centers of $u$ and $v$'s tiles of residence;
$f_u$ is the vertex--intrinsic social fitness~\cite{caldarelli2002scale} of vertex $u$, which controls $u$'s sociability, i.e., tendency to make friends.
The choice of $1$ as the distance dependence exponent, and of $f_u \sim 1+ \mathrm{Lognormal}\left(\ln(2),\frac{1}{4}\right)$ as used in the following is justified on an empirical basis in~\cite{Inferring} to best represent social contacts at the urban level.
A generalized formulation of the USN model used here is given in ~\cite{guarino2021model} and~\cite{Inferring}, where it is also shown that the main topological properties, s.a. the degree distribution, the clustering coefficient, and the average distance, are preserved in different network realizations. 
The implementation of the considered USN model is freely available as open-source software at \href{https://gitlab.com/cranic-group/usn}{gitlab.com/cranic-group/usn}.

We simulate the disease propagation using an individual--based continuous--time SIR model: at each time step, each infected individual $u$ spreads the disease to each susceptible individual $v$ with rate $p_{u,v} \beta$ and recovers with a fixed rate $\mu$.
If $\mathcal{S}_t$, $\mathcal{I}_t$, and $\mathcal{R}_t$ denote, respectively, the sets of susceptible, infected, and recovered individuals at time $t$, we assume that $|\mathcal{I}_0|=1$, i.e., a single individual, called \emph{index case}, is infectious at time 0.
The expected number of infections directly caused by the index case in a completely susceptible population is  $R^{\mathrm{index}}_0=\E\left[\sum_{v\neq u}\frac{p_{u,v}\beta}{\mu}\right]$, where $\E$ indicates the average w.r.t. the choice of the index case. 
As done in~\cite{biggerstaff2014estimates,liu2018measurability}, we calibrate the model on a common ILI (Influenza Like Illness) by setting the average time to recovery to 3 days and $R^{\mathrm{index}}_0=1.3$, thus obtaining $\mu=1/3$ and $\beta\approx0.0436$.

It may be useful to interpret $p_{u,v}$ as the probability that $u$ and $v$ come into contact and $\beta$ as the probability that a contact between an infectious agent and a susceptible one results in contagion.
Our epidemic model may thus be formulated in terms of a sequence of temporal contact networks $G_I^t=(V,E^t)$, where $E^t$ is the set of contacts occurring at time $t$ of the epidemic simulation, $(u,v)\in E^t$ with probability $p_{u,v}$, and, if $(u,v)\in E^t$, $u\in\mathcal{I}_t$ and $v\in\mathcal{S}_t$, the disease is transmitted from $u$ to $v$ with probability $\beta$.

\begin{table}[h]
\caption{The six configurations used for our epidemic model.}
\label{tab:configurations}
\centering
\scalebox{0.75}{
\begin{tabular}{p{45mm}p{20mm}p{17mm}p{17mm}p{60mm}}
\toprule
\multicolumn{2}{c}{\tabhead{Configuration}} & \multicolumn{3}{c}{\tabhead{$p_{u,v}$}} \\ 
\tabhead{Name} & \tabhead{Acronym}  & \tabhead{$(u,v)\in E_H$} & \tabhead{$(u,v)\in E_A$} & \tabhead{$(u,v)\notin E$} \\
\midrule
Homogeneous Mixing    & HM & $\propto |E|$ & $\propto |E|$ & $\propto |E|$ \\
Static Network        & SN & $1$ & $1$ & $0$ \\
Homogeneous Noise     & HN & $1$ & $0.5$ & $\propto |E|-|E_H|-|E_A|/2$ \\
Age--based Noise      & AN & $1$ & $0.5$ & $\propto s_{g_u,g_v}(|E|-|E_H|-|E_A|/2)$ \\
Distance--based Noise & DN & $1$ & $0.5$ & $\propto d(u,v)^{-1}(|E|-|E_H|-|E_A|/2)$ \\
Age-- and Distance--based Noise & ADN & $1$ & $0.5$ & $\propto s_{g(u),g(v)}d(u,v)^{-1}(|E|-|E_H|-|E_A|/2)$ \\
\bottomrule
\end{tabular}
}
\end{table}

We compare six configurations, summarized in Table~\ref{tab:configurations}.
In the HM configuration, each $G_I^t$ is an Erdős–Rényi graph, whereas in the SN configuration $G_I^t=G$ for all $t$.
In all other configurations, household relations induce a static frame of \emph{daily} contacts ($p_{u,v}=1$), acquaintance relations (e.g., between close friends or co-workers) induce \emph{frequent} contacts ($p_{u,v}=0.5$). In contrast, \emph{fortuitous} contacts (e.g., due to transportation, restaurants or stores) occur between agents not having any stable social relation.
The frequency of fortuitous contacts depends on: the age--based social mixing in AN, the geographic distance in DN,  both in ADN,  neither in HN.
By construction, all configurations satisfy $\sum_{u<v} p_{u,v} = |E|$, where $|E|$ is the number of edges of $G$\footnote{\cite{ball2008network} has no similar condition, but our noise is equivalent to theirs in the limit $N\to \infty$.}.
It follows that both the expected density of $G_I^t$ and the expected number of potential contagion events $\sum_{u<v} p_{u,v}\beta$ are fixed, thus making the six configurations comparable.

Our epidemic simulations are subject to statistical fluctuations, but the fluctuations induced by different network instances are negligible\footnote{These tests, omitted here due to space limitations, may be made available on request.}. We therefore ran 100 identical simulations on a fixed social network instance, so that all fluctuations are due to epidemic dynamics. 
With a single index case, the fraction of population involved in the outbreak follows a bimodal distribution.
The first of the two peaks corresponds to processes that die out in early stages of the epidemic, while the second corresponds to genuine outbreaks that involve a large fraction of the population.
The two peaks are well separated at a fraction of around $25\%$ of the population involved so that we can easily discard simulations resulting in early extinction.
In the following sections, we report the average results over the $\approx30$ experiments actually giving rise to an epidemic outbreak.

\section{Epidemic Threshold, Evolution and Reproduction Number}

By simulating the spread of an infectious disease, we may understand how susceptible a territory or population is to the occurrence of an epidemic outbreak.
This is usually done by looking at:
(i) the evolution of the number of infected and recovered individuals at time $t$, respectively denoted $I_t$ and $R_t$;
(ii) the epidemic threshold $\beta_c$, defined as the critical value for the transmission parameter $\beta$ after which the fraction of recovered individuals is asymptotically finite;
(iii) the epidemic reproduction number $R(t)$, which measures the expected number of new infections caused by a single individual that becomes contagious at time $t$, during the entire course of his/her infectiousness period;
(iv) the basic epidemic reproduction number $R_0^{Index}$, corresponding to the expected number of infections caused by the index case in an entirely susceptible population.
All of these quantities are clearly related to the contact patterns, which, in our case, are described by the degree distribution of the temporal network of contacts $G_I^t$.

\begin{figure}
    \centering
    \begin{subfigure}{.32\textwidth}
        \centering
        \includegraphics[width=\textwidth]{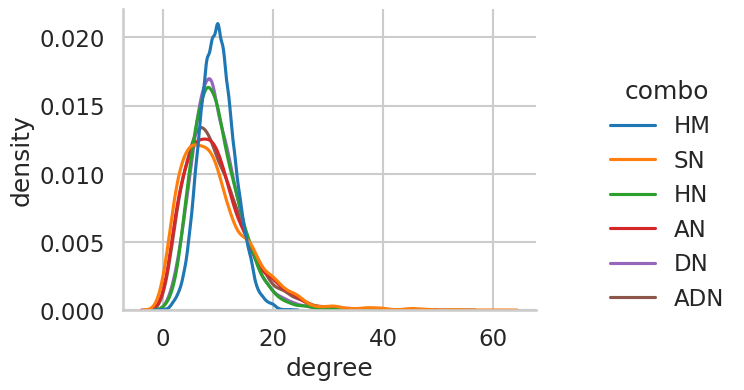}
        \vspace*{-5ex}
        \caption{}
    \label{fig:degree}
    \end{subfigure} 
    \hfill
    \begin{subfigure}{.32\textwidth}
        \centering
        \includegraphics[width=\textwidth]{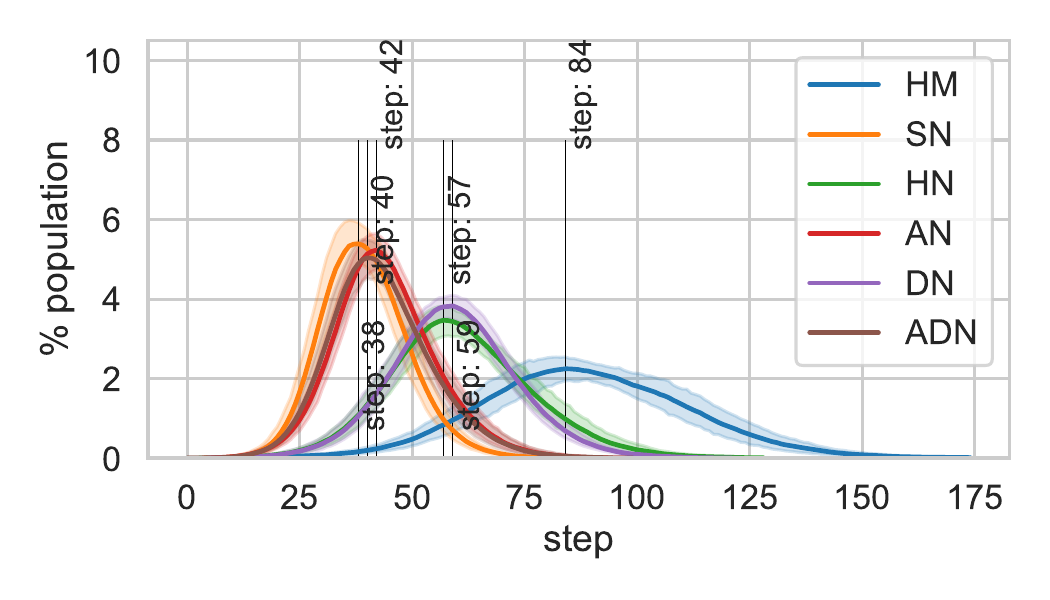}
        \vspace*{-5ex}
        \caption{}
        \label{fig:infected}
    \end{subfigure}  
    \hfill
    \begin{subfigure}{.32\textwidth}
        \centering   \includegraphics[width=\textwidth]{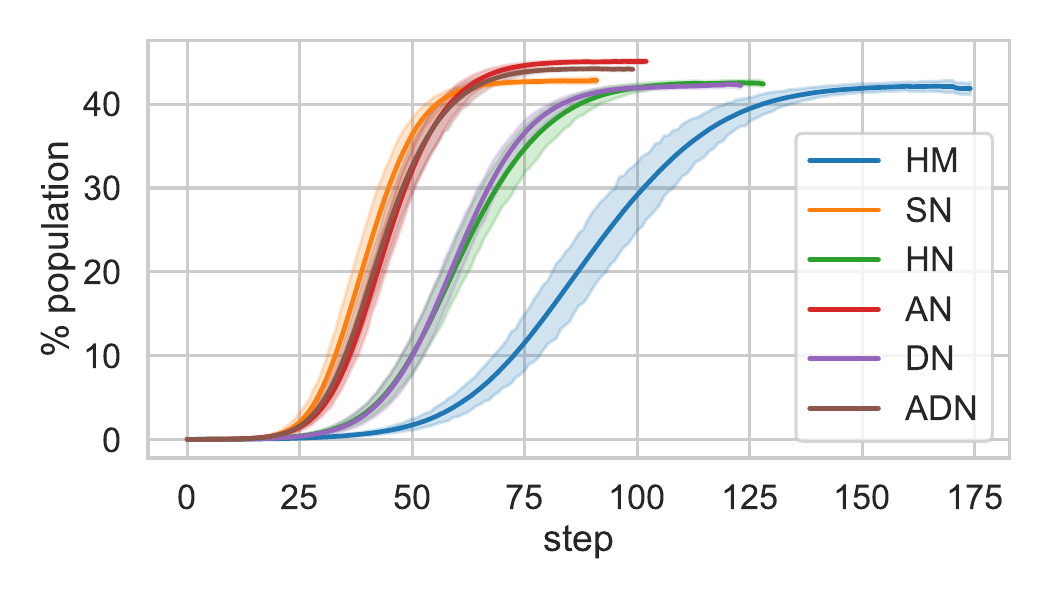}
        \vspace*{-5ex}
        \caption{}
        \label{fig:recovered}
    \end{subfigure} 
    
    \begin{subfigure}{0.32\textwidth}
        \centering
        \includegraphics[width=\textwidth]{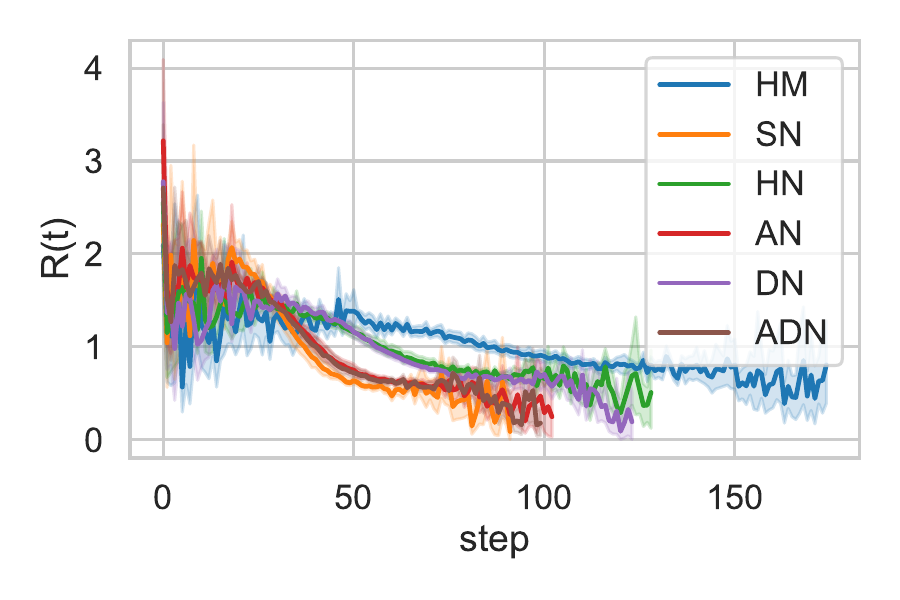}
        \vspace*{-5ex}
        \caption{}
        \label{fig:Rt}
    \end{subfigure}
    \hfill
    \begin{subfigure}{0.32\textwidth}
        \centering
        \includegraphics[width=\textwidth]{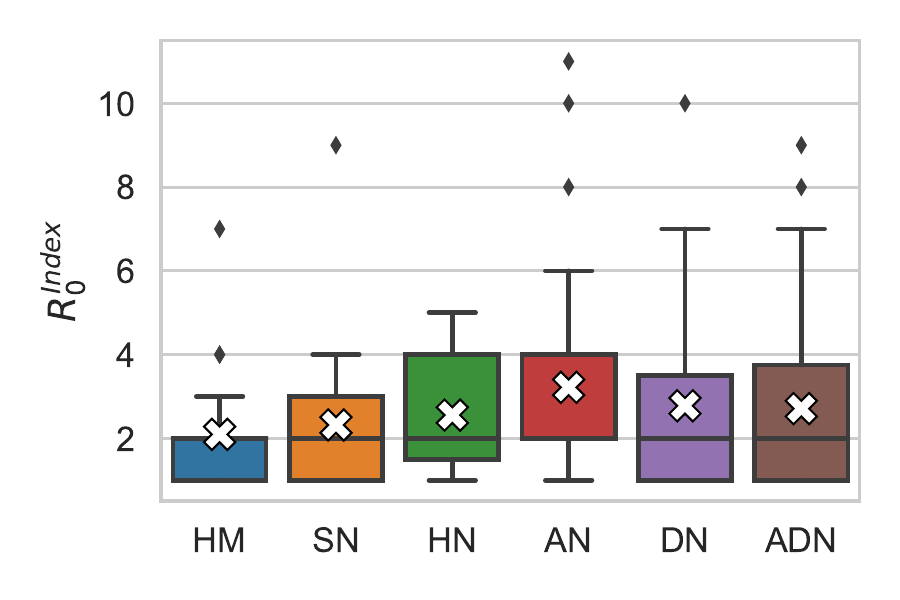}
        \vspace*{-5ex}
        \caption{}
        \label{fig:R0_dist}
    \end{subfigure}
    \hfill
    \begin{subfigure}{.32\textwidth}
        \centering   \includegraphics[width=\textwidth]{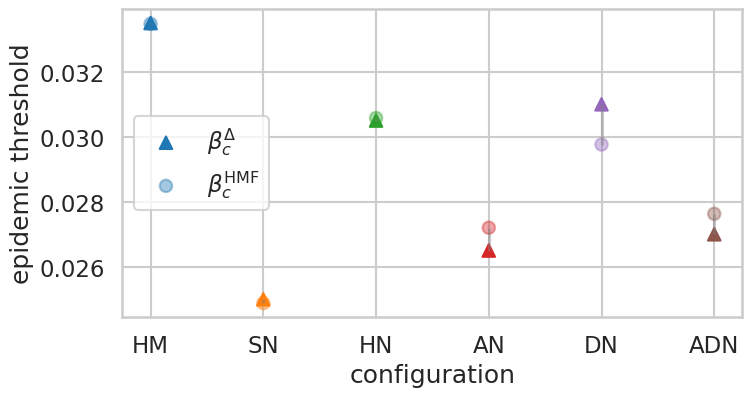}
        \vspace*{-5ex}
        \caption{}
        \label{fig:threshold_comparison}
    \end{subfigure}
    \caption{For all six configurations: (a) degree distribution of the temporal networks $G_I^t$; (b,c) evolution of the number of infected ($I_t$) and recovered ($R_t$) individuals; (d) evolution of the reproduction number $R(t)$; (e) distribution of the empirical basic reproduction number; (f) empirical epidemic threshold.}
    \label{fig:SIR}
\end{figure}

By comparing the degree distribution of $G_I^t$ (Fig.~\ref{fig:degree}) and the evolution of $I_t$ and $R_t$ (Figs.~\ref{fig:infected} and~\ref{fig:recovered}), we immediately notice three clusters:
(i) under the assumption of homogeneous mixing (HM configuration) the degrees follow a Poisson distribution and the epidemic spread is slower and overall less pervasive than in any data-driven and more realistic configuration;
(ii) considering a combination of recurrent contacts, induced by the static social network, and of age--homogeneous fortuitous contacts, whether distance--based or not (HN and DN), produces a skewed degree distribution that causes a sharp acceleration of the outbreak and a slight increase of the total number of cases;
(iii) if instead, all contacts are governed by age--based preferences (SN, AN, and ADN), the degree distribution has a longer right tail, and we observe an even earlier peak in the number of infected and an even greater attack rate.

$R(t)$ provides another way to look at the evolution of the epidemic, especially in view of the implementation of potential mitigation strategies.
Fig.~\ref{fig:Rt} confirms what already emerged in Figs.~\ref{fig:infected} and~\ref{fig:recovered}: the evolution of the epidemic is almost entirely controlled by the degree distribution of the temporal network of contacts, with the amount and nature of fortuitous contacts playing a major role, but with the dependence of these contacts on distance being almost irrelevant.
Fig.~\ref{fig:R0_dist} shows that $R_0^{Index}$ is generally greater in data-driven and realistic configurations, characterized by heterogeneous patterns of contacts.
Since we considered only simulations in which an epidemic occurred
($>25\%$ of recovered, as discussed in Section~\ref{sec:model}), Fig.~\ref{fig:R0_dist} highlights how the emergence of an epidemic outbreak is generally associated to the index case being a socially active individual.

The analytical computation of the epidemic threshold for the SIR model on networks is generally based either on the heterogeneous mean--field (HMF) theory, or on the analogy with the bond percolation, leading to expressions for $\beta_c$ that depend on the first and second moment of the degree distribution.
In~\cite{ball2008network} the authors estimate $\beta_c$ in a population with two levels of mixing approximating the initial stages of the epidemic with a two--types branching process.
To the best of our knowledge, however, no analytical expression is known for our general formulation with multiple levels of mixing -- nor for the equivalent formulation in terms of temporal networks.
We determined the threshold empirically as in~\cite{shu2015numerical}, i.e., as $\beta_c^{\Delta} = \arg \max \Delta(\beta)$, where $\Delta(\beta)=\frac{\sqrt{\mean{\rho^2}-\mean{\rho}^2}}{\mean{\rho}}$ is the epidemic variability and $\rho=\rho(\beta)$ is the attack rate. 
Only for the two extreme cases, the HM and SN configurations, the HMF approximation $\beta_c^{\mathrm{HMF}}=\frac{\avgdeg}{\mean{\mathrm{deg}^2} - \avgdeg}$ is theoretically valid -- where $\avgdeg$ and $\mean{\mathrm{deg}^2}$ are computed based on the degree distribution of the temporal networks $G_I^t$.
We found, respectively, $\beta_c^{\mathrm{HMF}}(\mathrm{HM}) \approx 0.0335$ and $\beta_c^{\mathrm{HMF}}(\mathrm{SN}) = 0.0249$, in perfect agreement with $\beta_c^{\Delta}$.
Not surprisingly, the entirely static network SN and the network with HM lie at the two ends of the spectrum, raising doubts on their adequacy to describe realistic contact models.
The other four configurations only differ for the way fortuitous contacts are drawn: imposing a dependence on age makes the system more vulnerable to epidemic outbreaks, while assuming a dependence on distance makes it slightly more resistant.

\begin{figure}
    \centering
    \begin{subfigure}{.32\textwidth}
        \centering
        \includegraphics[width=\textwidth]{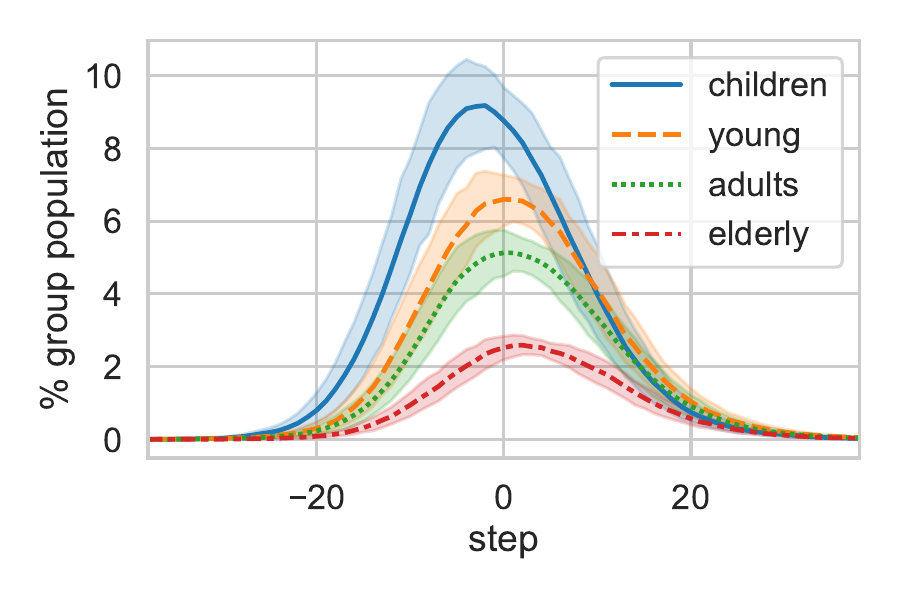}
        \includegraphics[width=\textwidth]{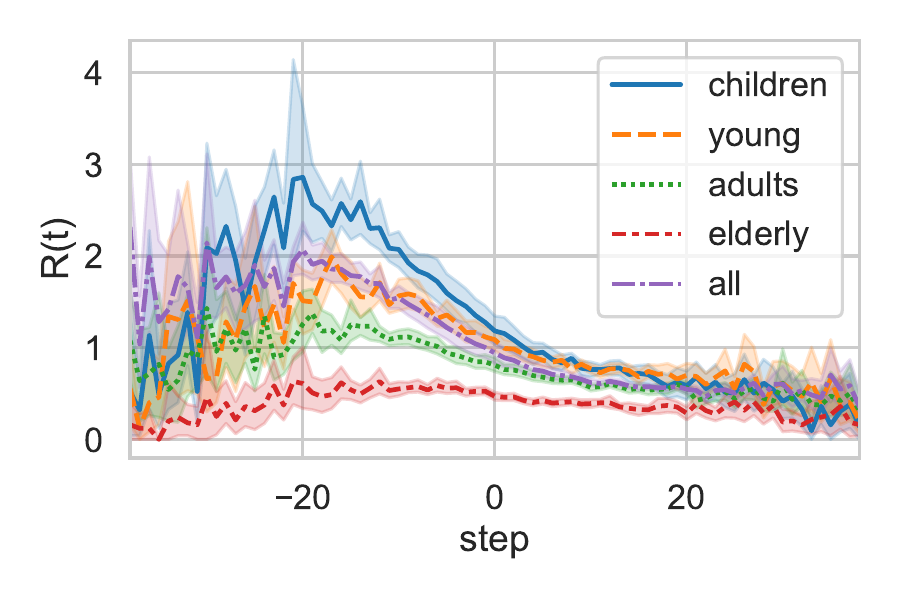}
        \vspace*{-5ex}
        \caption{SN.}
        \label{fig:age_SN}
    \end{subfigure} 
    \hfill
    \begin{subfigure}{.32\textwidth}
        \centering
        \includegraphics[width=\textwidth]{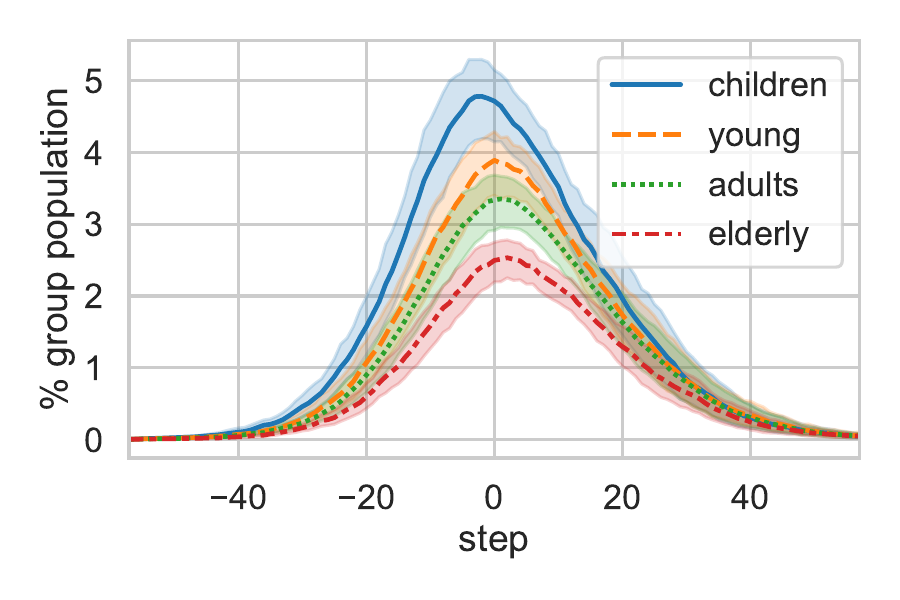}
        \includegraphics[width=\textwidth]{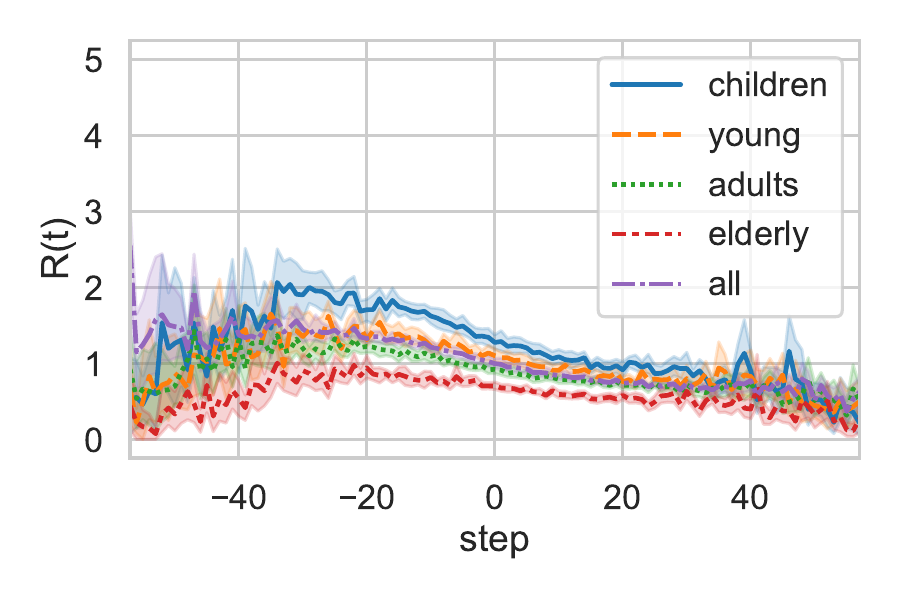}
        \vspace*{-5ex}
        \caption{HN.}
        \label{fig:age_HN}
    \end{subfigure} 
    \hfill
    \begin{subfigure}{.32\textwidth}
        \centering
        \includegraphics[width=\textwidth]{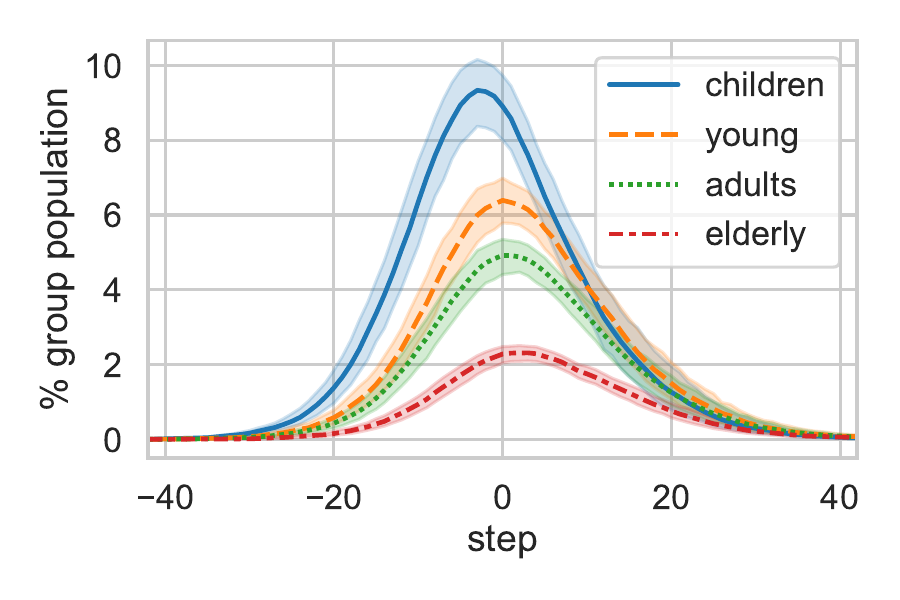}
        \includegraphics[width=\textwidth]{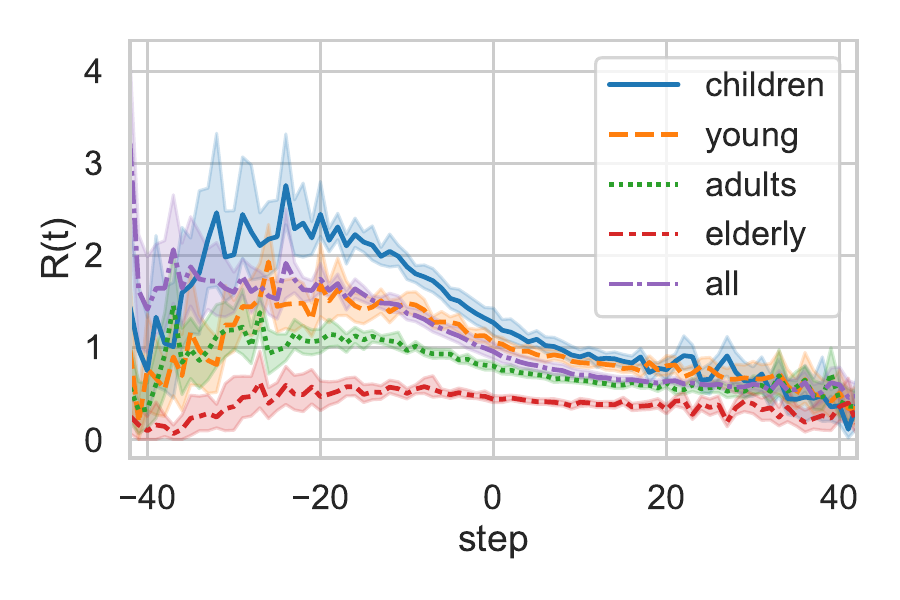}
        \vspace*{-5ex}
        \caption{AN.}
        \label{fig:age_AN}
    \end{subfigure} 
    
    \caption{For the three models SN (a), HN (b) and AN (c): evolution of the number of infected individuals in each age-group and of the epidemic reproduction number for each age-group.}
    \label{fig:age}
\end{figure}

To gain a better understanding of the importance of an age-stratified population, Fig.~\ref{fig:age} shows the time evolution of the number of infected individuals $I_t$ and of the reproduction number $R(t)$ for the four age--groups, shifted with respect to the global epidemic peak, for the SN, HN and AN models (HM is trivial, DN and ADN are similar to HN and AN, respectively). 
Children and young adults seem to drive the epidemic and to be more impacted, as expected given their greater average degree and internal cohesion.
The fact that the SN and AN configurations are almost indistinguishable suggests that the diffusion of the disease in different age-groups is dominated by the social mixing, with the recurrence of contacts playing a minor role.

\section{Geographic Spreading and Predictability}

\begin{figure}
    \centering
    \begin{subfigure}{.32\textwidth}
        \centering
        \includegraphics[width=\textwidth]{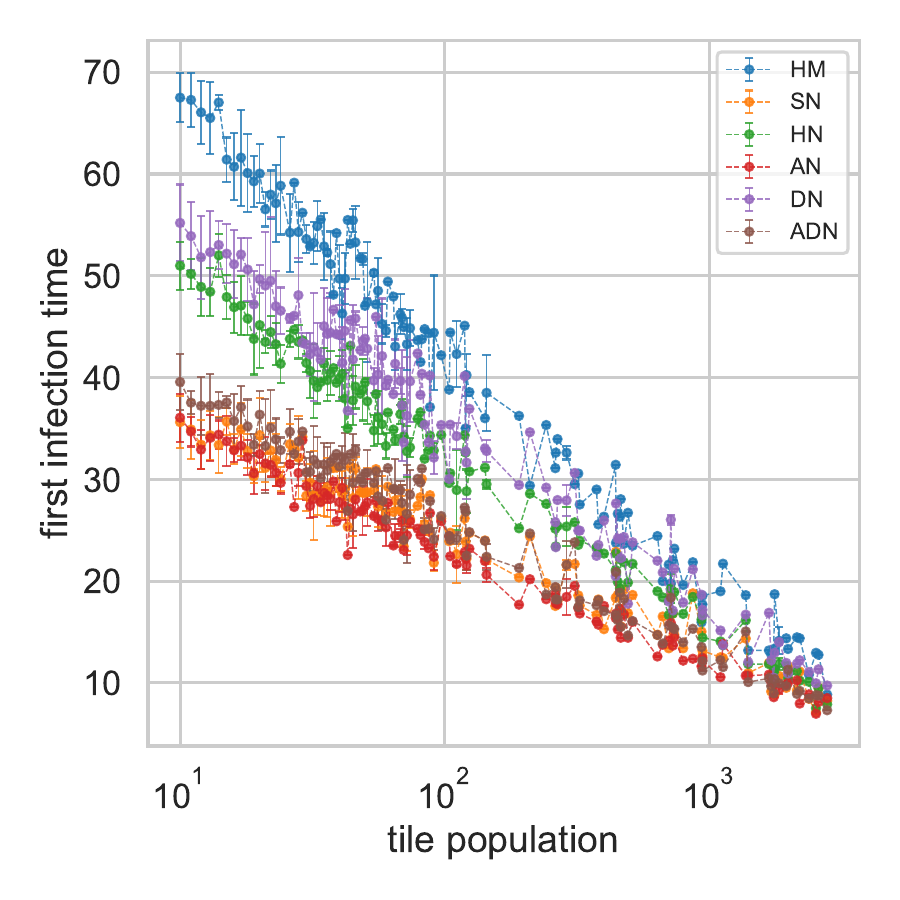}
        \vspace*{-5ex}
        \caption{}
        \label{fig:first_infection}
    \end{subfigure}
    \hfill
    \begin{subfigure}{.32\textwidth}
        \centering
        \includegraphics[width=\textwidth]{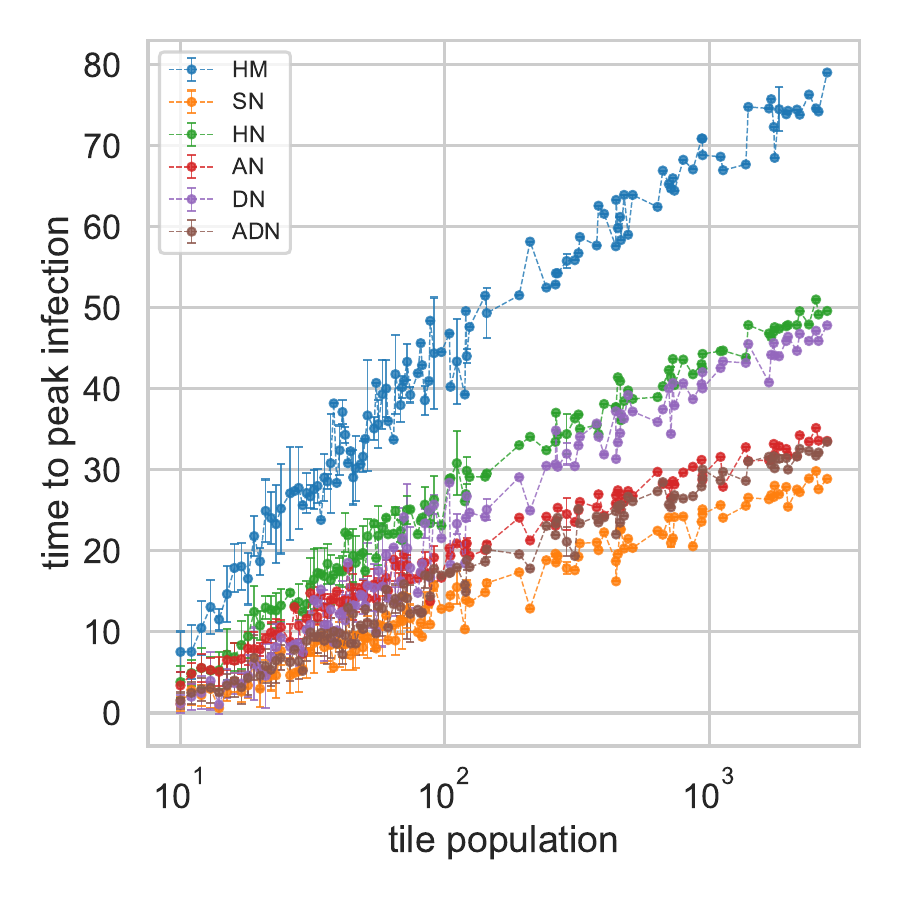}
        \vspace*{-5ex}
        \caption{}
        \label{fig:peak_infection}
    \end{subfigure}
    \hfill
    \begin{subfigure}{.32\textwidth}
        \centering
        \includegraphics[width=\textwidth]{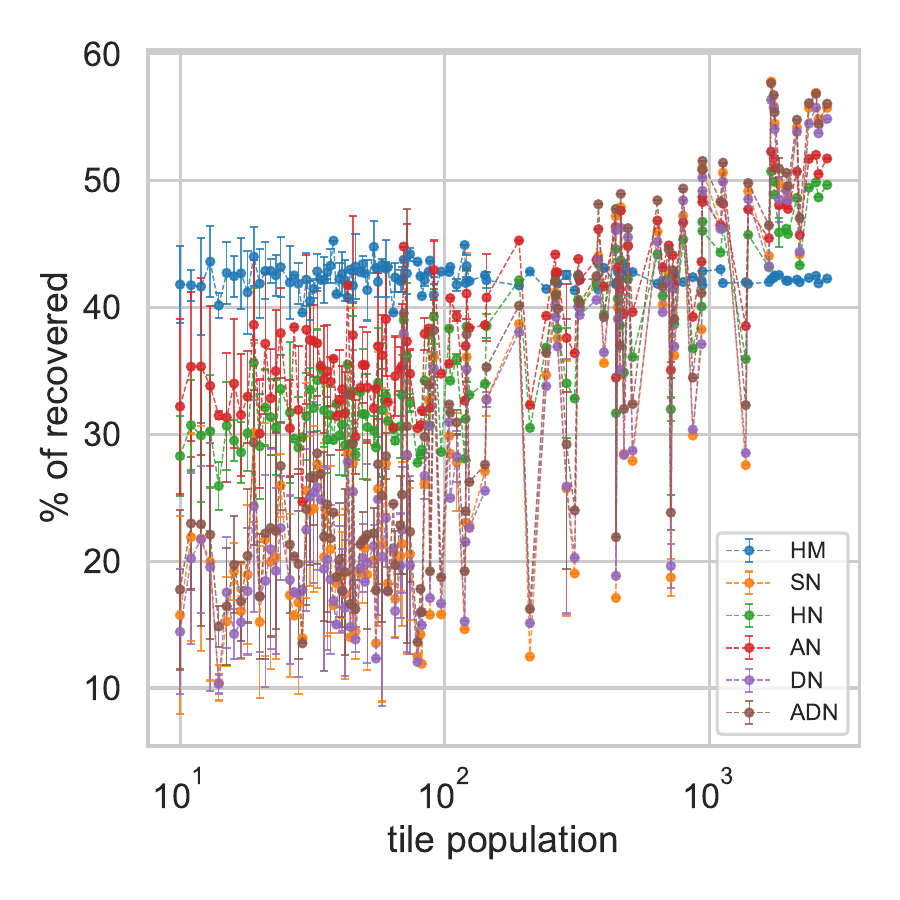}
        \vspace*{-5ex}
        \caption{}
        \label{fig:tot_recovered}
    \end{subfigure}
    \caption{For each model configuration, as a function of the tile population: (a) mean time of first infection in the tile; (b) mean time interval between first infection and epidemic peak; (c) attack rate. Averages are across equally populated tiles.}
    \label{fig:pop_count}
\end{figure}

A primary goal of data--informed epidemic models is gaining insights into the mechanisms driving the geographic spread of the disease.
To assess the role of population density, in Fig.~\ref{fig:pop_count} we show the time of the first infection in a tile, the time interval between the contagion entering a tile and the epidemic peak within the tile, and the attack rate (i.e., fraction of recovered at the end of the epidemic), all as a function of the tile population (or, equivalently, the density, our tiles covering a fixed area).
The time of the first infection (Fig.~\ref{fig:first_infection}) decreases linearly with the logarithm of the population.
The infection reaches densely populated tiles in 10 to 15 days, regardless of the configuration, whereas for scarcely populated tiles different ways to model the contact patterns yield very different estimates for the time of the first infection and, hence, for the available time span by which any mitigating actions aimed at preserving low--density areas must be put in place.
Both the time to reach the peak of infections (Fig.~\ref{fig:peak_infection}) and, more notably, the attack rate (Fig.~\ref{fig:tot_recovered}, except for the HM configuration for which the tile of residence is irrelevant) seem to present two regimes: a  sub--logarithmic time to the peak and a constant attack rate for tiles with less than $\approx100$ inhabitants; an approximately logarithmically increasing time to the peak and attack rate for densely populated tiles.
This behavior -- which requires further investigation -- may be related to the different ratio of inter-- and intra--tile infections in areas with different population density,
which may also explain why Fig.~\ref{fig:tot_recovered} is the only case in which the SN, DN and ADN configurations, united by the fact that contacts with individuals living in other tiles are less likely, show an analogous trend: a generally lower attack rate in scarcely populated tiles, compensated by a greater attack rate in high--density tiles.

To understand how the disease travels in the city and whether it follows predictable paths, we focus on the time evolution of three quantities: the fraction of infected tiles, the normalized entropy of the local prevalence vector and the overlap function measuring the similarity between two different outbreak realizations~\cite{colizza2006role}.
Formally, let: 
$N_j$ be the number of inhabitants of tile $j$;
$I_j(t)$ be the number of infected individuals in tile $j$ at time $t$; 
$\vec{q}(t)$ be the normalized vector of prevalence of infected individuals in each tile, with components $q_j(t)=\frac{I_j(t)}{N_j\sum_l I_l(t)/N_l}$; 
$i(t)=\sum_j\frac{I_j(t)}{N}$ be the total epidemic prevalence at time $t$ and $\vec{i}(t)=(i(t),1-i(t))$; 
$\vec{\pi}(t)$ be the vector of probabilities $\pi_j(t) = \frac{I_j(t)}{\sum_l I_l}$ that an infected individual is in tile $j$ at time $t$;
$\theta^{a,b}(t)=\mathrm{sim}_H(\vec{i}^a(t), \vec{i}^b(t)) \cdot \mathrm{sim}_H(\vec{\pi}^a(t), \vec{\pi}^b(t))$ be the overlap between two different outbreak realizations $a$ and $b$, where $\mathrm{sim}_H(\vec{x}, \vec{y}) = \sum_j \sqrt{x_j y_j}$ is the Hellinger affinity.
The fraction of infected tiles, defined as $\tau(t)=\frac{|\{j:I_j(t)>0\}|}{T}$, measures the geographic prevalence of the infection at time $t$.
The normalized entropy $H(t)=-\frac{1}{\log T}\sum_j q_j(t)\log q_j(t)$ of $\vec{q}(t)$ is a measure of the level of geographic heterogeneity of the disease prevalence (and, hence, of the riskiness) at time $t$. 
The predictability of the system at time $t$, empirically computed as the distribution of the overlap $\theta^{a,b}(t)$ over all pairs of simulations $a,b$, measures the similarity between different realizations.

\begin{figure}
\centering
    \begin{subfigure}{.32\textwidth}
        \centering
        \includegraphics[width=\textwidth]{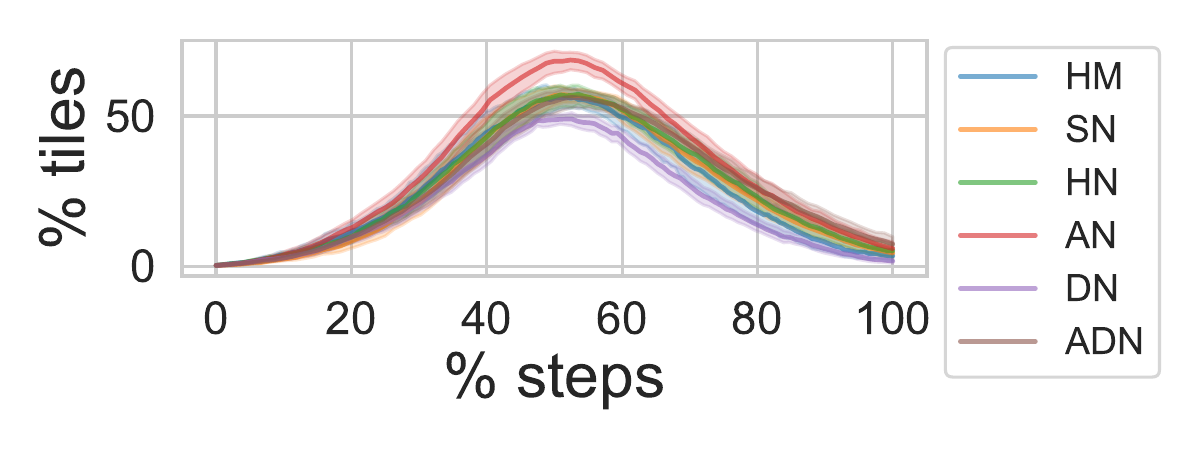}
        \includegraphics[width=\textwidth]{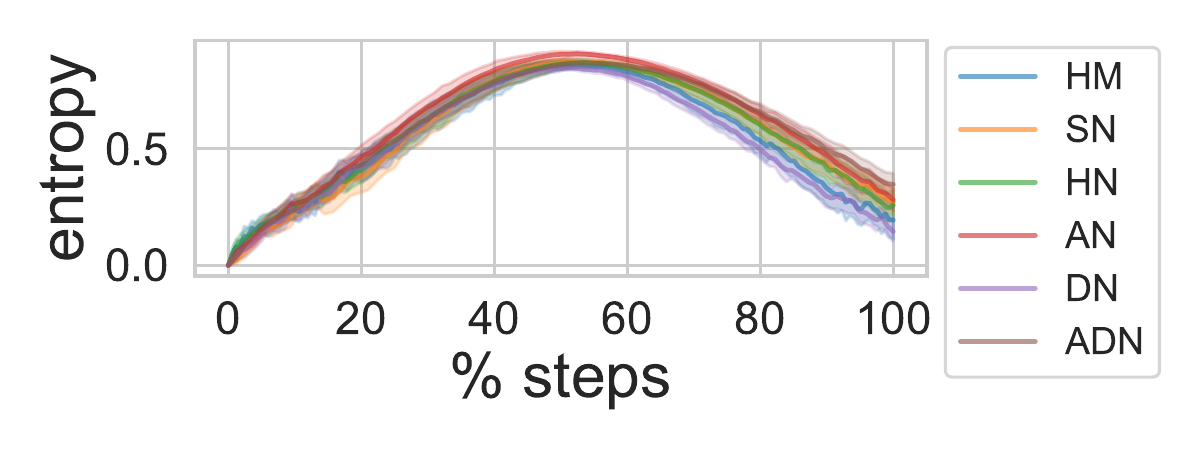}
        \includegraphics[width=\textwidth]{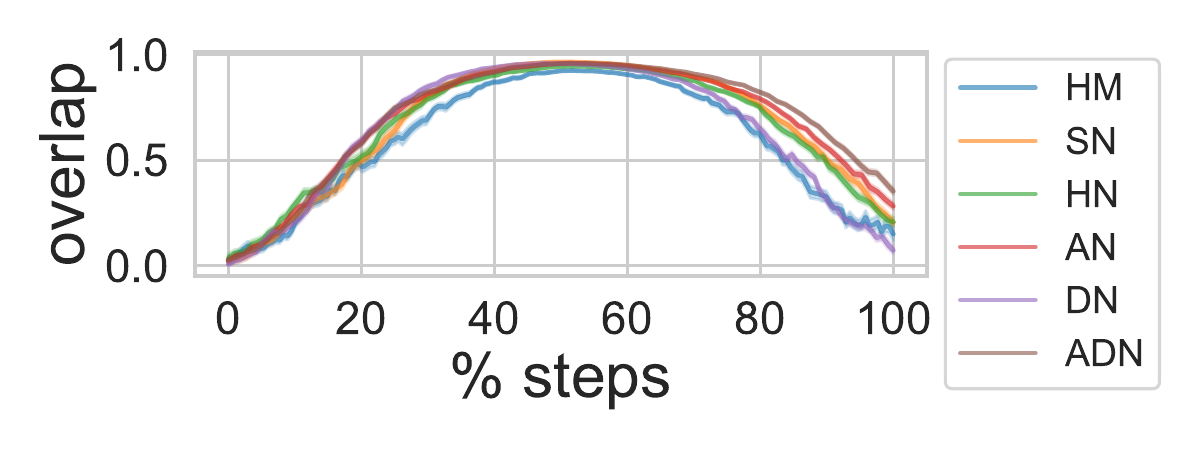}
        \vspace*{-5ex}
        \caption{Start from random tile.}
        \label{fig:predictability_entropy_random}
    \end{subfigure}
    \hfill
    \begin{subfigure}{.32\textwidth}
        \centering
        \includegraphics[width=\textwidth]{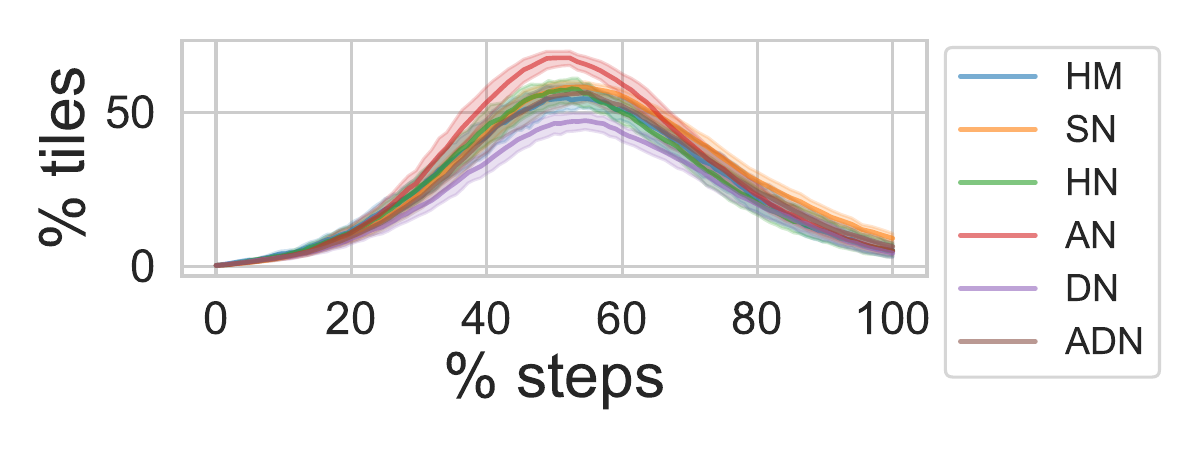}
        \includegraphics[width=\textwidth]{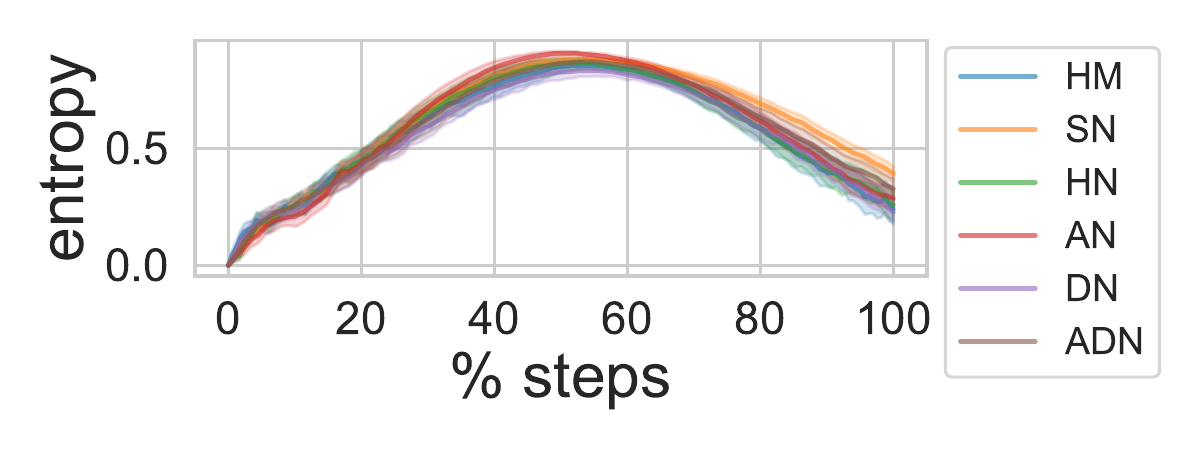}
        \includegraphics[width=\textwidth]{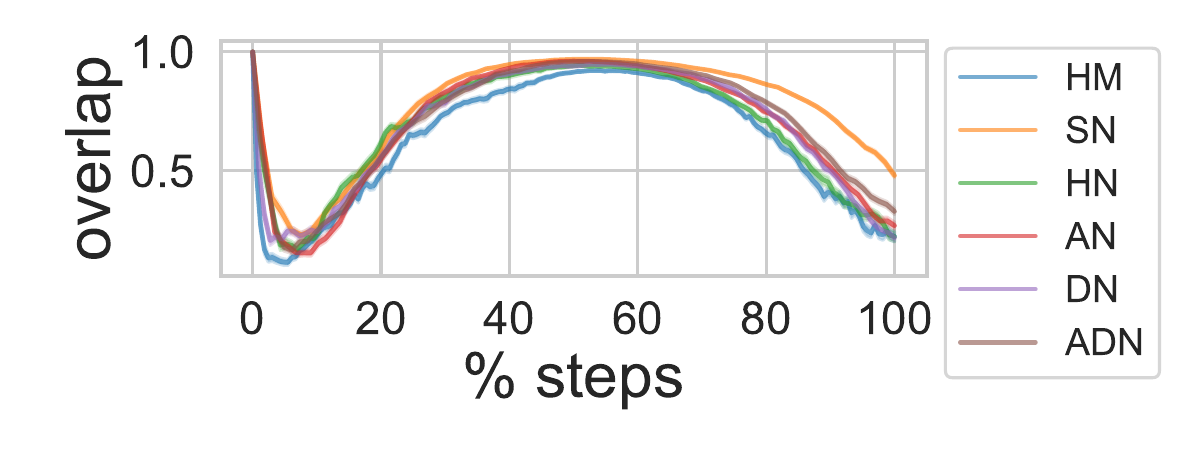}
        \vspace*{-5ex}
        \caption{Start from central tile.}
        \label{fig:predictability_entropy_central}
    \end{subfigure}
    \hfill
    \begin{subfigure}{.32\textwidth}
        \centering
        \includegraphics[width=\textwidth]{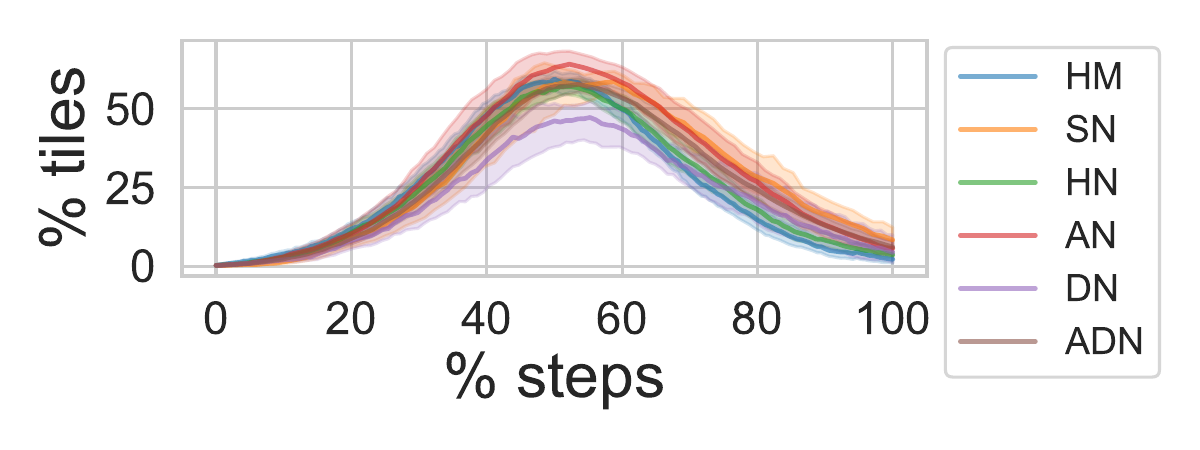}
        \includegraphics[width=\textwidth]{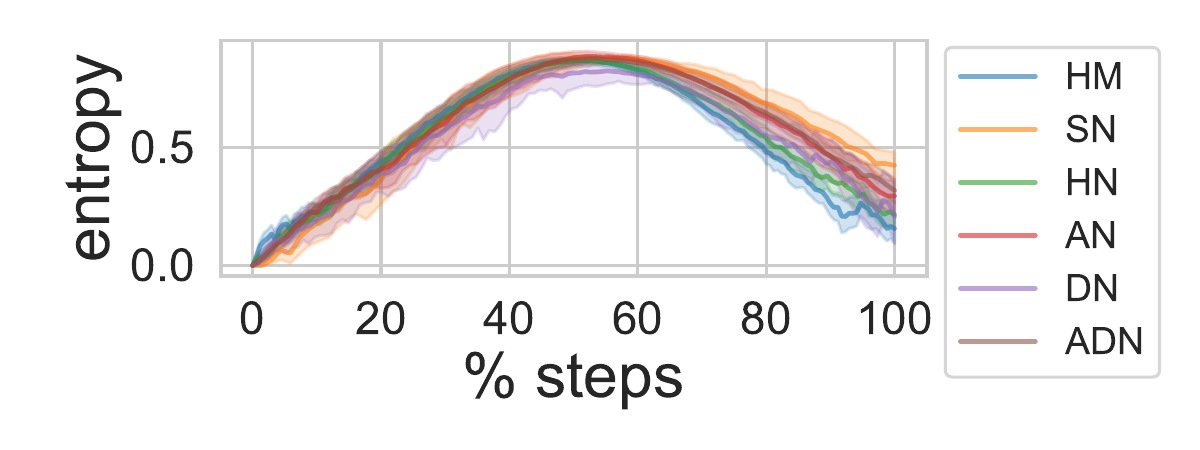}
        \includegraphics[width=\textwidth]{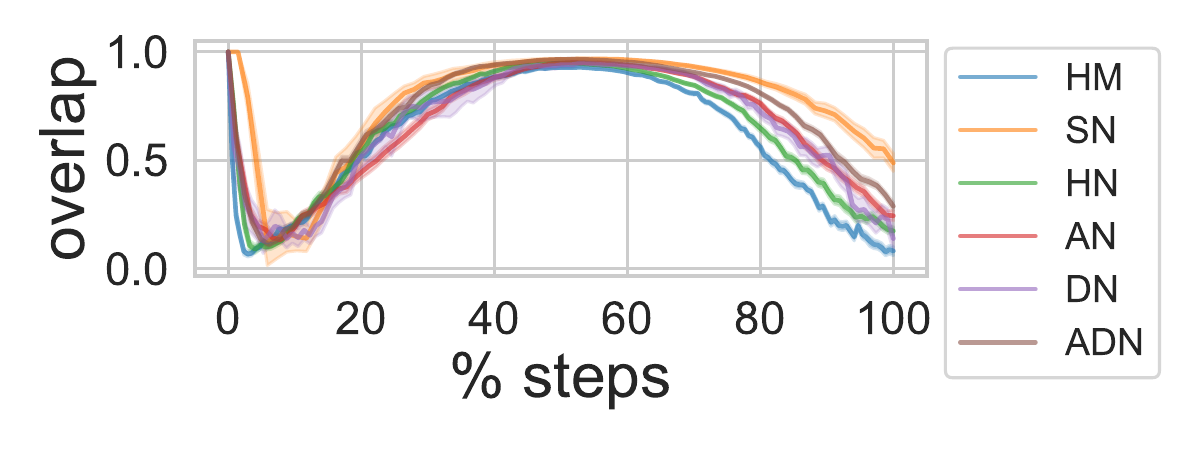}
        \vspace*{-5ex}
        \caption{Start peripheral tile.}
        \label{fig:predictability_entropy_peripheral}
    \end{subfigure}
    \caption{For all six configurations, fraction of infected tiles (top), normalized entropy of the local prevalence (center) and predictability (bottom) and when the index case is chosen: (a) at random in the entire population; (b) at random in a central tile; (c) at random in a peripheral tile. The time scale is normalized to make all curves comparable.}
    \label{fig:predictability_entropy}
\end{figure}

In Fig.~\ref{fig:predictability_entropy} we plot $\tau(t)$, $H(t)$ and $\Theta^{a,b}(t)$ (average with 95\% confidence interval over $a,b$) for epidemic simulations in which the index case is chosen uniformly at random in the entire population (Fig.~\ref{fig:predictability_entropy_random}), in a central tile (Fig.~\ref{fig:predictability_entropy_central}) or in a peripheral tile (Fig.~\ref{fig:predictability_entropy_peripheral}).
To discount the overall speed of propagation of the infection and make the curves directly comparable, the time scale has been normalized.
The fraction of infected tiles is always $<0.7$ and the AN and DN configurations differ from all others for having, respectively, a visibly greater and lower value of $\tau(t)$ around the epidemic peak.
This means that fortuitous contacts tend to favor or contrast the geographic spread of the infection based on whether such contacts are age-- or distance--dependent.
The normalized entropy never reaches 1 and is significantly $<1$ for most of the time in all configurations, meaning that the epidemic behavior is geographically heterogeneous and some tiles are, broadly speaking, more dangerous than others.
The evolution of the epidemic is only predictable in the proximity of the epidemic peak, which is also when $\tau(t)$ and $H(t)$ are maximal, i.e., when the infection is already highly prevalent in most of the city.
At the same time, the fact that the overlap is large close to the peak tells us that we may try to predict which areas are somehow more dangerous at the peak.
Even when the tile of origin of the infection is fixed (Figs.~\ref{fig:predictability_entropy_central} and~\ref{fig:predictability_entropy_peripheral}), the predictability drops very quickly at the beginning of the epidemic, to then follow a similar trend to the case where the index case is chosen uniformly at random (Fig.~\ref{fig:predictability_entropy_random}).
The behavior of $H(t)$ and $\Theta^{a,b}(t)$ does not differ appreciably among configurations, except that the system is generally more homogeneous and predictable with the increase of the constraints.

\section{Conclusion}
\label{sec:conclusions}
To contribute to the understanding of epidemic outbreaks in an urban territory, we defined an individual--based SIR model for a data--driven synthetic population of agents connected by a social network of stable social relations.
Our model assumes that an individual has daily contacts with her household members, frequent contacts with a small network of acquaintances (e.g., friends or coworkers), and fortuitous contacts with the rest of the population.
We considered four configurations that differ for the dependence of fortuitous contacts on demographic and/or geographic features of the host population, along with two benchmark configurations respectively characterized by the assumption of a random network (homogeneous mixing), or a static contact network.

We found that age--structure in contact patterns causes a right--skewed degree distribution that makes the epidemic faster and more pervasive.
Imposing that the interaction frequency decays with distance as $d^{-1}$, as suggested by previous empirical studies, seems to have little effect on the high--level diffusion patterns.
A faster decay may result in a more significant impact~\cite{medo2020contact}, but it would require empirical justification.
The distance still plays a role in determining the frequency of inter--tile infections, which affects the relation between the population of a tile and its attack rate.
Population density has a considerable impact on the local incidence of the epidemic and its temporal evolution.
In particular, the attack rate shows two distinct behaviors for low/high density populated tiles. Our result echoes the empirical finding of two different diffusion regimes on the country/city scale with low/high population density \cite{grenfell2001travelling,viboud2006synchrony}.
This analogy could indicate a pattern of hierarchical spatial diffusion even in urban areas when population heterogeneity is taken into account, but further investigation is needed.

We noticed the absence of backbones  capable of defining preferred epidemic pathways.
This fact is not entirely surprising since all our configurations present heterogeneous connectivity patterns and a wide range of alternative and equivalent travel routes for the infection.
In data--driven meta--population models, such as the one used in~\cite{colizza2006role} to describe the role of the airline transportation network in global epidemics, the local and global dynamics of contagion are well separated, and, at the broader scale, the high number of possible spreading channels is compensated by the presence of dominant connections in traffic flow.
This is not the case for our model of urban contacts, where the only stratification of contacts is induced by the discretization of the distance function $d(u,v)$ used to create the acquaintance network, and, possibly, to establish the likelihood of fortuitous contacts.
Understanding whether the absence of dominant epidemic pathways is inherently due to the different scale of the analysis or it can be addressed by informing the model with additional data (e.g., mobility, transportation, phone calls) is a primary direction for future work. 

\bibliographystyle{unsrt}
\bibliography{biblio}

\end{document}